# Ultralow and Tunable Thermal Conductivity of Parylene C for Thermal Insulation in Advanced Packaging

Yicheng Wei[1,+], Han Xu[2,+], Xingqiang Zhang[2], Wei Wang[1,2,*], Zhe Cheng[1,2,*]

[1] School of Software & Microelectronics, Peking University, Beijing 100871, China

[2] School of Integrated Circuits and Beijing Advanced Innovation Center for Integrated Circuits, Peking University, Beijing 100871, China

[+]These authors contributed equally

[*]Authors to whom correspondence should be addressed: w.wang@pku.edu.cn; zhe.cheng@pku.edu.cn

**Abstract**

Parylene C thin films have significant applications in advanced packaging of microelectronics. Their thermal properties are critical for thermal management of electronic devices. However, a unified understanding of the tunable structure and the corresponding thermal conductivity is still missing. This study investigated parylene C thin films of varying thickness and post-annealing temperatures (200 °C and 320 °C) grown via thermal chemical vapor deposition. The ultralow thermal conductivity of as-deposited parylene C measured by time-domain thermoreflectance (TDTR) is 0.10-0.13W/m-K. The thermal conductivity can be tuned by post-annealing. Significant increase in thermal conductivity is observed in the samples annealed at 320 °C (0.18-0.24W/m-K) which induces melting and recrystallization. The results of XRD and polarized Raman spectroscopy show that the enhanced thermal conductivity is due to improved crystalline quality and the change in chain orientations. The measured thermal conductivities of the as-deposited and 200 °C-annealed films are much lower than the values predicted by the Cahill's minimum thermal conductivity model, which can be explained by the diffuson-mediated minimum thermal conductivity model. Parylene C is found to possess the lowest thermal conductivity among dense low-k materials. Our work provides guidance for the structural design of ultra-low thermal conductivity polymers and corresponding thermal design of electronics.

# 1. Introduction

Parylene C, a semi-crystalline polymer renowned for its exceptional dielectric properties, chemical stability, and ability to form ultra-thin conformal films, has become an important functional coating in microelectronics[1–4]. In critical applications—ranging from insulating encapsulation for complementary metal-oxide-semiconductor (CMOS) devices to structural protection for micro-electro-mechanical systems (MEMS)[5–7], its linear-chain-based structure provides not only robust mechanical toughness but also vital electrical insulation[8,9]. Moreover, in three-dimensional integrated circuits (3D-IC), both thermal insulation and low dielectric constant are important for electronics packaging. However, when deposited as thin films, parylene C often exhibits structural and thermal properties that deviate significantly from its bulk counterparts. Despite its widespread applications, the complexity of its nanoscale film structure has left a gap in the comprehensive understanding of thermal transport, particularly regarding how the chain structure governs thermal conductivity.

Fundamentally, thermal transport in polymers originates from atomic vibrations propagating through the molecular network, which is bonded by covalent bonds and *van der Waals* bonds, contributing to the intra-chain and inter-chain heat transport, respectively[10]. The thermal resistance of a strong covalent bond is approximately an order of magnitude lower than that of a *van der Waals* bond[11]. Consequently, the primary bottleneck for thermal transport in polymers lies in the inefficient inter-chain heat transfer due to the weak *van der Waals* forces. Bulk polymers typically exhibit low thermal conductivities due to disordered structure and entanglements of molecular chains, but the thermal conductivity of a single polymer chain is several orders of magnitude higher because of less scatterings[12] along the chain axis originating from the stiff covalent bonding[13]. Particularly, semi-crystalline polymers comprise amorphous and crystalline parts. In semi-crystalline polymers, this process is bifurcated: in crystalline regions, thermal conductivity is dominated by phonons, constrained by phonon-phonon and phonon-boundary scatterings; in amorphous regions, vibrational states are categorized as locons, diffusons, and phonon-like propagons[14–17]. While locons are localized and contribute minimally to heat transfer[18], the extended modes (diffusons and propagons) serve as the primary carriers. The bonding strength of the structure partially determines the extent to which each extended vibrational mode contribute to thermal conductivity[16]. Particularly in thin films, the mean free path of these propagons is often limited by the size effect, leading to a reduction in cross-plane thermal conductivity[15,16,19].

As a representative semi-crystalline polymer, parylene C possesses a complex structure governed by multiple structural factors, including crystallinity, crystallite size, inter-chain spacing, and molecular chain orientations[20,21]. Kachroudi[21] and Kim[22] *et al.* indicate that annealing at temperatures below the melting point will increase crystallinity rate from 45%–60% to 75%–85% and crystallite size from 4 nm-6 nm to 8 nm-16 nm. The measured thermal conductivities of parylene C with thicknesses of 210 nm, 440 nm, and 760 nm were 0.95, 0.99, and 0.99 W/m-K, respectively, showing very slight thickness dependence[23]. It is also found that, due to surface confinement, parylene C exhibits an amorphous structure at thicknesses below 40 nm, while crystallinity and crystallite size increase as thickness increases[24]. When the crystallite size is less than the phonon mean free paths in the crystalline region, boundary scatterings become dominant and reduce thermal conductivity[15]. The controls of crystallinity and chain orientations are typically achieved via process modulation, including post-deposition annealing and deposition temperature tuning during the thermal chemical vapor deposition (CVD) process [25]. The systematic research on the relations between crystalline quality, molecular chain orientations, and thermal conduction mechanisms in parylene C is lacking.

With the miniaturization of integrated circuits, the distance between high-power and temperature-sensitive components decreases, making unexcepted heat crosstalk and degrading device and system performance. For example, Advanced 2.5D packaging brings logic and memory closer to facilitate near-memory computing [26]. However, the high heat flux from high-power logic dies poses a critical threat to the adjacent temperature-sensitive memory. To balance the requirements of high-speed communication and thermal reliability, effective thermal insulation strategies must be implemented with the package to hinder heat transport[27]. Due to the ultralow thermal conductivity of parylene C, it is important to study the heat transfer mechanisms and thermal conductivity modulation methods for thermal insulation.

This work prepared parylene C films of varying thicknesses via thermal CVD and investigated their structural and thermal transport properties, as well as the effects of the post-annealing process. X-ray diffraction (XRD) and polarized Raman spectroscopy were employed to characterize crystalline quality, including crystallinity, inter-chain spacing, and crystallite size, as well as molecular chain orientations. The picosecond acoustic technique and time-domain thermoreflectance (TDTR) technique were used to measure sound velocities and thermal conductivities, respectively. By correlating film thickness, annealing temperature, and phonon transport behavior, this study reveals ultralow and tunable thermal conduction in parylene C films, which sheds light on the heat transfer mechanism in ultralow thermal conductivity polymers.

## 2. Methods

The parylene C thin films, with target thicknesses of 100 nm, 300 nm and 1 μm, were deposited on 10 mm×10 mm highly doped silicon sheets using a standardized PDS 2010 thermal CVD system, utilizing 0.16 g, 0.48 g, and 1.6 g of parylene C dimers, respectively. The deposition parameters included a pyrolysis temperature of 690 °C, a base pressure of 6 mTorr, and a deposition pressure of 15 mTorr. Post-deposition, the films underwent thermal annealing under a nitrogen atmosphere at temperatures of 200 °C or 320 °C for a total duration of 2.5 hours (30 minutes ramp-up and 2 hours steady-state hold). Subsequent thickness measurements using a profilometer revealed a thickness reduction in the films annealed at 320 °C, which is attributed to the melting of the polymer's amorphous regions.

To characterize the crystallite size, crystallinity, and inter-chain spacing of the parylene C thin films, XRD measurements were conducted using a Bruker D8 Advance system. Samples were analyzed in their as-deposited state and after being subjected to annealing at 200 °C and 320 °C. A CuKα radiation source (λ = 1.5418Å) was employed, operating at 40 kV and 40 mA. The XRD patterns were recorded over a 2θ range of 10°–30° with a step size of 0.04°.

Polarized Raman spectroscopy (Horiba/LabRam HR Evolution) was used to analyze the molecular chain orientations of parylene C films. A 532 nm solid-state laser was used as the excitation source. The spectra were collected with a resolution of 1 cm$^{-1}$ over the range of 580–1800 cm$^{-1}$, focusing on the characteristic peak at ~1608 cm$^{-1}$ (assigned to benzene ring C-C stretching vibration, sensitive to chain orientation) and ~695 cm$^{-1}$ (assigned to C-Cl stretching vibration). Measurements were conducted under orthogonal polarization configurations. The film was fixed on a rotatable stage and Raman spectra were recorded every 10° over a polarization angle range of 0°–180°.

Time-domain thermoreflectance (TDTR)[28–30] is a pump-probe technology, used to measure the cross-plane thermal conductivity of parylene C thin films. A pump laser heats the sample, while a delayed probe laser monitors the temperature change and is collected via thermoreflectance. The collected signal, which varies across samples and embodies their thermal properties, is fitted to a multilayer heat transfer model[31] to infer unknown thermal parameters such as thermal conductivity. The uncertainty of the TDTR system is estimated to be ±8%. A 5× objective was used, and the pump and probe spot diameter is 22.2 μm and 9.9 μm. All measurements are conducted at a modulation frequency of 5.68 MHz.

## 3. Results and Discussions

### 3.1. Tunable structure

The thermal CVD and forming process of parylene C films are shown in Figure 1. In this study, nine samples were prepared and categorized into three groups: as-deposited parylene C with thicknesses of 90, 260, and 970 nm; samples annealed at 200 °C with thicknesses of 90, 257, and 960 nm; and those annealed at 320 °C with thicknesses of 50, 160, and 680 nm. Notably, the glass transition point is ~90 °C, and the melting point is ~290 °C[25]. Therefore, annealing at 320 °C involves melt-recrystallization.

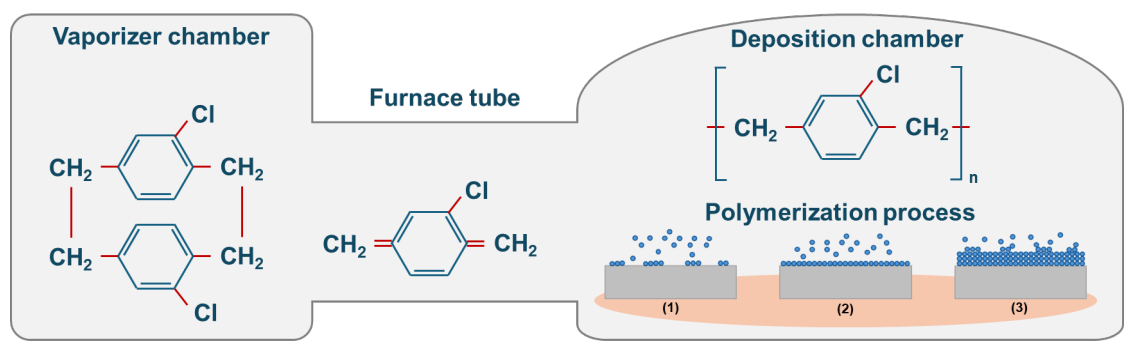

Figure 1. The forming process of parylene C thin films via thermal CVD.

Figure 2 shows the XRD and polarized Raman results of all samples. All the XRD patterns are smoothed and subtracted from the baseline, as the baseline is consistent across all samples. The displayed polarized Raman characterizations correspond to the peak located at ~1608 cm$^{-1}$. The XRD patterns without subtraction of the baseline and the polarized Raman characterizations regarding the ~695 cm$^{-1}$ peak of parylene C are shown in the supplementary information. The diffraction peak at ~14° corresponds to the parylene C (020) plane within the monoclinic unit cell[21,24,32]. Meanwhile, the polarization direction of the laser in polarized Raman spectroscopy varies in a plane parallel to the sample, so parylene C molecular chains align along the c-axis and stack in the b-axis direction[33], namely, molecular chains are parallel to the surface[32].

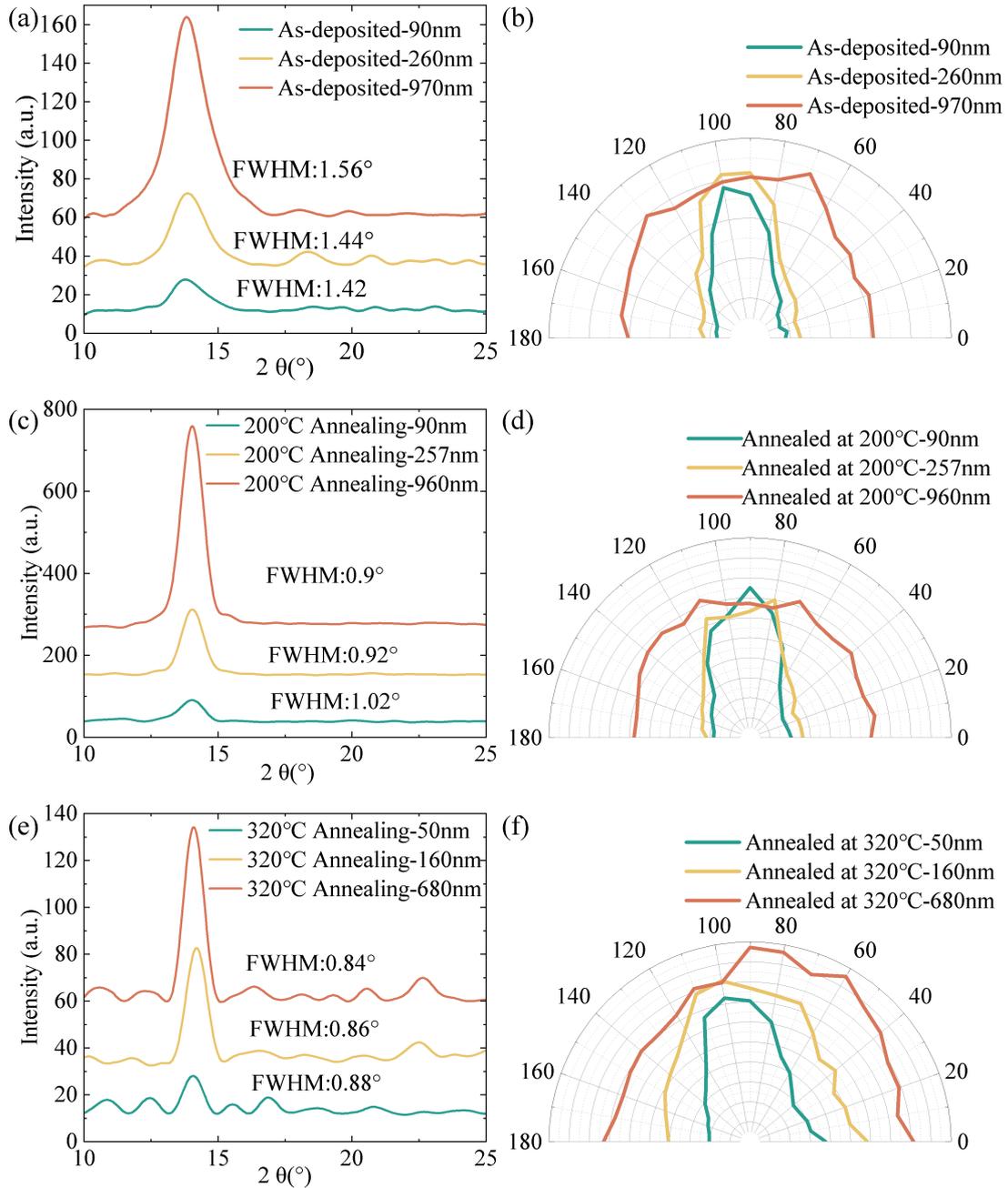

Figure 2. The XRD and polarized Raman characterizations of all samples. (a) The XRD patterns and (b) polarized Raman spectra of the three as-deposited parylene C thin films with different thicknesses. (c) The XRD patterns and (d) polarized Raman spectra of the three parylene C thin films annealed at 200 °C. (e) The XRD patterns and (f) polarized Raman spectra of the three parylene C thin films annealed at 320 °C.

The XRD patterns in Figure 2 reveal that the annealing results in a narrower full width at half maximum (FWHM), indicating larger crystallite sizes or reduced lattice distortions within the structure. Particularly, annealing at 200 °C results in a significant increase in diffraction peak intensity of the XRD signal, indicating enhanced crystallinity and crystalline quality, which is consistent with the reported work[21]. This effect is more pronounced in thicker Parylene C films. For parylene C annealed at 320 °C, although the intensity of the (020) peak shows no significant change relative to the as-deposited state, the FWHM is approximately reduced to half of that of the as-deposited parylene C, meaning the average crystallite size becomes roughly doubled based on the Debye–Scherrer formula. During the melt-recrystallization process, molecular chains are freed from the "frozen" state imposed by surface confinement during deposition. The original unstable microcrystals melt and rearrange to seek the lowest energy configuration. This results in larger crystalline regions that are more stable and have much less distortion. This combination of narrower peaks (larger crystalline regions) but unchanged intensity suggests a mechanism of melt-recrystallization. In addition, with the annealing temperature or thickness increasing, the systematic shift of the (020) diffraction peak toward higher angles indicates a reduction in the inter-chain spacing according to the Bragg relation, which can be observed more clearly in the supplementary information. This densification of the structure effectively strengthens the inter-chain *van der Waals* interactions, which facilitates efficient heat conduction.

The polarized Raman spectra in Figure 2 reveal a clear thickness-dependent orientation across all sample groups. In thinner films such as 90 nm, the pronounced anisotropy in the polar plots suggests that surface confinement forces the molecular chains into a preferential in-plane alignment with a specific direction. However, as the film thickness increases toward the micrometer scale, the curves in polar coordinates become increasingly circular, indicating a shift from surface-confined anisotropy to bulk-like isotropy, where molecular chain orientation approaches a random distribution.

In addition, the polar coordinate diagrams of 320 °C-annealed parylene C exhibit a more circular (isotropic) shape compared to the as-deposited and 200 °C-annealed samples. Annealing at 200 °C facilitated minor in-situ rearrangements of the molecular chains. In contrast, annealing at 320 °C triggered a phase transition into the molten state, providing the molecular chains with sufficient energy for complete relaxation. Consequently, the melt-recrystallization process effectively eliminates the preferential orientation, yielding a highly isotropic spatial distribution. Although the overall order decreases, some molecular chains tilt or even become perpendicular, thereby facilitating efficient phonon transport along the cross-plane direction through strong covalent bonds rather than the weak *van der Waals* bonds.

*3.2. Sound velocity measurements*

Picosecond acoustic technique is used to measure the sound velocity[34,35], which shares an experimental setup with TDTR showed in Figure 3(a). An ultrafast pump pulse (a femtosecond laser) heats the sample surface, generating a stress pulse. The stress wave propagates into the sample until it encounters an interface. A portion of the stress wave continues propagating deep into the sample, while another portion reflects back to the sample surface and changes the sample's optical reflectivity, detected by a delayed probe pulse. Consequently, the intensity of the reflected probe signal is altered[36]. Figures 3(b-c) show the picosecond acoustic echoes of the three samples. As for the samples annealed at 320 °C, echo1 is located at 27 ps, and echo2 is located at 158 ps. Combined with the thickness of the parylene C films, the longitudinal sound velocity can be determined (160 nm/(158 ps-27 ps)/2). Figure 3(d) shows the longitudinal sound velocities of five samples. As the annealing temperature increases, the longitudinal sound velocity increases, which facilitates phonon transport.

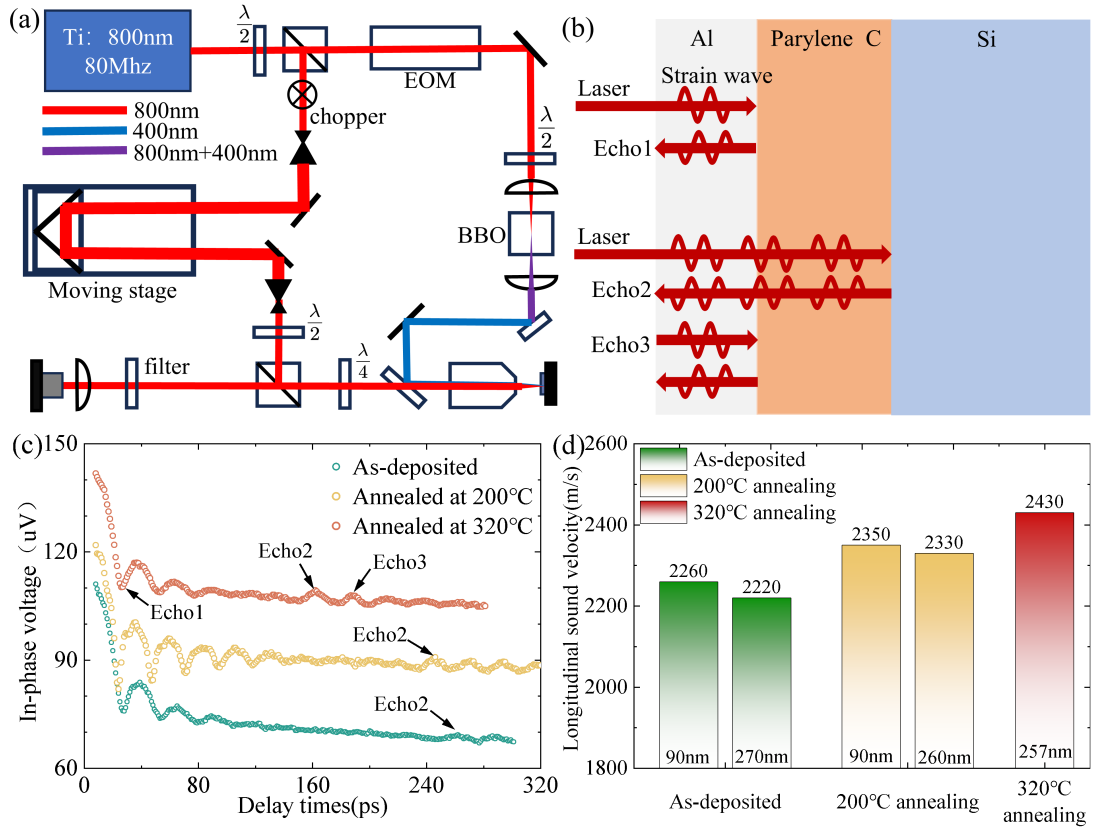

Figure 3. (a) The schematic diagram of the TDTR experimental setup. (b) The schematic diagram of the picosecond acoustic mechanism in samples. (c) Picosecond acoustic echoes of the three samples with different thicknesses. The green circles indicate that parylene C has a thickness of 260 nm. The yellow circles correspond to a thickness of 257 nm, while the red ones correspond to a thickness of 160 nm. (d) Longitudinal phonon velocities of the five samples measured by the picosecond acoustic technique.

*3.3. The ultralow and tunable thermal conductivities*

TDTR was used to measure the thermal conductivity of the parylene C thin films. Before measurements, an ~80-nm-thick Al layer was deposited on the samples as a transducer. The sensitivities of TDTR signals to the thermal parameters of the parylene C samples are shown in Figure 4(a). The high sensitivity of the thermal conductivity means it can be measured precisely. The experimental data is fitted with the solution of a multilayer heat transfer model of the sample structure. For the parameters in the data fitting, the thermal conductivity of the Al transducer is measured by the four-point probe method combined with the Wiedemann-Franz law. The thermal conductivity of the highly doped silicon substrate is measured in advance as an input parameter in the data fitting. The heat capacity of Al and Si substrate are from the previously reported works[37–39]. Figure 4(b) shows an example of the data fitting of TDTR measurements.

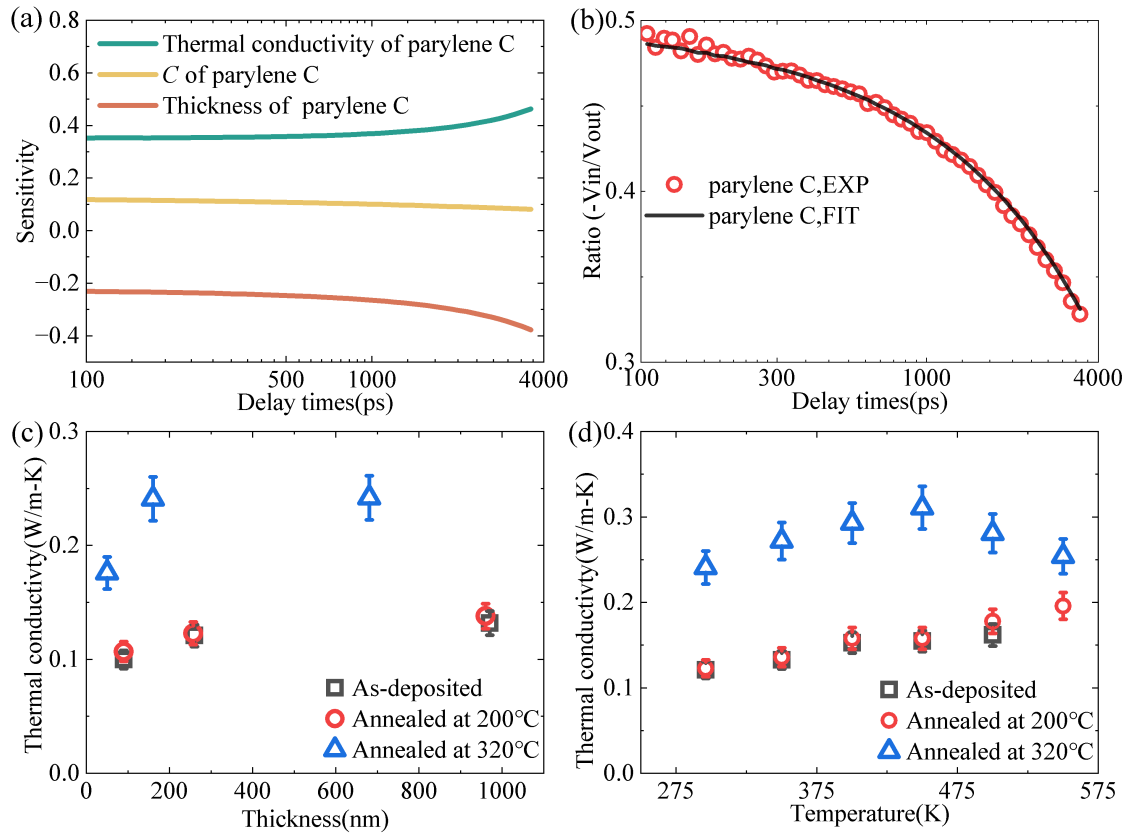

Figure 4. (a) The sensitivities of TDTR signals to the thermal conductivity, the heat capacity, and the thickness of the parylene C annealed at 200 with a 90 nm thickness. (b) The TDTR data fitting curve: the red circles represent experimental data, while the black line represents the fitting data. (c) The thickness-dependent thermal conductivity of the parylene C thin films with different annealing conditions. (d) The measured thermal conductivities of the parylene C films as a function of temperature. Blue triangles, black squares, and red circles represent the 260-nm-thick as-deposited parylene C thin film, the 257-nm-thick parylene C annealed at 200 °C, and the 160-nm-thick parylene C annealed at 320 °C, respectively.

Figure 4(c) presents the measured thermal conductivities of the total nine samples. Specifically, the thermal conductivities of the as-deposited parylene C are 0.10 W/m-K, 0.12 W/m-K, and 0.13 W/m-K, while the 200°C-annealed parylene C exhibits slightly higher values of 0.11 W/m-K, 0.12 W/m-K, and 0.14 W/m-K, respectively, depending on the corresponding thickness. The thermal conductivity of the parylene C thin films is tunable via post-annealing. Under the same annealing conditions, the thermal conductivity increases with film thickness, exhibiting a clean size effect. When the thickness exceeds the mean free paths of phonons, the intensity of phonon-boundary scattering is reduced. Beyond this traditional size effect, the transformation of molecular chain orientation emerges as a critical factor. In ultra-thin films, a preferential in-plane chain orientation severely hinders phonon transport in the cross-plane direction due to the weak inter-chain *van der Waals* interactions. In thicker films, the weakening of surface confinement allows for a more randomized molecular chain arrangement, introducing more vertical covalent pathways that facilitate cross-plane heat conduction.

Consistent with the aforementioned XRD and polarized Raman characterizations, annealing at 320 °C, which corresponds to the melt-recrystallization regime, significantly enhances thermal conductivity (reaching 0.18 W/m-K, 0.24 W/m-K and 0.24 W/m-K depending on different thicknesses). This enhancement is driven by a synergistic mechanism: the improvement in crystal quality-characterized by enlarged crystallite sizes and tighter chain spacing-extends the phonon mean free path and strengthens inter-chain coupling. Simultaneously, the transition of structural isotropy reorients a fraction of the high-thermal-conductivity covalent backbones toward the cross-plane direction. While the former reduces phonon scattering, the latter provides an effective pathway for vertical phonon transport, collectively overcoming the bottleneck of cross-plane inter-chain thermal conduction. In summary, the melt-recrystallization induced by 320 °C annealing serves as a critical pathway for optimizing the crystalline structure and enhancing the cross-plane thermal conductivity of parylene C.

However, the thermal conductivity of the parylene C annealed at 200 °C remains largely unchanged compared to the as-deposited parylene C, for both ambient or elevated temperatures, although the aforementioned crystallinity and longitudinal sound velocity increase. This suggests that the inter-chain *van der Waals* force is the primary bottleneck for phonon transport, as the parylene C annealed at 200 °C maintains a similar structural anisotropy to the as-deposited state. The enhancement of crystal quality is insufficient to overcome the dominant thermal resistance arising from weak inter-chain coupling in the cross-plane direction. Consequently, a significant enhancement in cross-plane thermal conductivity is only realized at 320 °C, where the melt-recrystallization process mitigates this bottleneck by reorienting covalent pathways toward the cross-plane direction.

Figure 4(d) shows the temperature-dependent thermal conductivity of the 260-nm-thick as-deposited parylene C, the 257-nm-thick parylene C annealed at 200 °C, and the 160-nm-thick parylene C annealed at 320 °C, respectively. Samples are heated using HCP421V hot plate of Instec company from 296 K to 600 K at 50 °C intervals and measured by TDTR at the same time. The thermal conductivities of the as-deposited parylene C and the annealed parylene C at 200 °C are close and gradually increase with temperature, while the thermal conductivity of the annealed parylene C at 320 °C rises first, and then decreases, which is similar to the polymer of 50% crystalline structure displayed in the reported work[40]. The heat capacity of parylene C at elevated temperatures used in the TDTR data fitting is set as a constant 0.918 MJ/m$^3$-K[23] based Debye theory.

$$\Theta_D = \frac{\hbar}{\kappa_B}(6\pi^2 \frac{N}{V})^{\frac{1}{3}}v = \frac{\hbar}{\kappa_B}(6\pi^2 \frac{n_{\text{eff}}\rho NA}{M})^{\frac{1}{3}}v \qquad (1)$$

$$\frac{3}{v^3} = \frac{1}{v_L^3} + \frac{2}{v_T^3} \qquad (2)$$

$$v_L = \left[\frac{E(1-\sigma)}{(1+\sigma)(1-2\sigma)\rho}\right]^{1/2} = \left[\frac{2(1-\sigma)}{(1-2\sigma)}\right]^{1/2} v_T \tag{3}$$

where $\hbar$ is the reduced Planck constant; $\kappa_B$ is the Boltzmann constant; $N/V$ is the number of atoms per unit volume; $NA$ is Avogadro constant; $M$ is the molar mass of parylene C calculated to be 138.59 using repeating unit chemical formula $C_8H_7Cl$; $n_{eff}$ is effective atomic number of the repeating unit; $\rho$ is the density with a value of 1289 kg/m³[23,39]; $\sigma$ is the Poisson's ratio with a value of 0.4[41]; $v$ is the average speed of sound; $v_L$ is longitudinal sound velocity measured by the picosecond acoustic technique and displaced in Figure 3(d), and $v_T$ is the transverse sound velocity.

Because hydrogen (H) atoms are not thermally excited at room temperature[42], they are not included in $n_{eff}$. Simultaneously, C and Cl atoms all lose 0.5 vibrational modes for each bond, and the benzene ring corresponds to two equivalent atoms[43,44]. The $n_{eff}$ can be calculated to 25/6, which is directly related to heat conduction[45]. Using the Equations (1-3), the transverse phonon velocity $v_T$ can be calculated to 992 m/s when longitudinal phonon velocity is taken as 2430 m/s, and the Debye temperature of parylene C is calculated to be 96 K, meaning that the atomic vibrations of parylene C are all excited. Therefore, it is feasible to assume that the heat capacity at high temperatures is identical to that at room temperature.

*3.4. Heat transfer mechanisms*

To figure out the heat transfer mechanism, the "minimum thermal conductivity" (also called the amorphous limit of thermal conductivity) of parylene C was calculated using Cahill's minimum thermal conductivity model, assuming scattering occurs at a distance equal to the interatomic spacing[46–48]：

$$\Lambda_{min} = \left(\frac{\pi}{6}\right)^{1/3} k_B n^{2/3} \sum_{i=1}^{3} v_i \left(\frac{T}{\Theta_i}\right)^2 \int_0^{\Theta_i/T} \frac{x^3 \exp(x)}{(\exp(x)-1)^2} dx \qquad (4)$$

**Where** $n$ is the number of atoms per unit volume; $v_i$ is the sound velocity in the different polarization directions, namely one longitudinal and two transversal, and the meanings of the other parameters are the same as those mentioned above.

In our experiments where $T \gg \Theta$, Equation (4) can be simplified as[17,48]:

$$\Lambda_{min} = \left(\frac{\pi}{48}\right)^{1/3} k_B n^{2/3} (v_L + 2v_T) \qquad (5)$$

The calculated minimum thermal conductivity using Equation (5) shown in Figure 5 is ~0.19 W/m-K. This value is higher than the measured thermal conductivities of the as-deposited and 200 °C-annealed samples, but remains lower than that of the parylene C annealed at 320 °C. The enhanced thermal conductivity of the parylene C annealed at 320 °C is attributed to its high crystalline quality and the efficient cross-plane covalent pathways, whereas the minimum thermal conductivity corresponds to the amorphous limit. When the temperature exceeds 500 K, the thermal conductivities start to reduce, gradually approaching the amorphous limit. This is partly because, as approaching the melting point, the material becomes increasingly disordered. Another contributing factor is the enhanced phonon-phonon scatterings.

The values calculated by the minimum thermal conductivity model exceed the measured values for the as-deposited parylene C and annealed parylene C at 200 °C. Agne *et al.* proposed a diffuson-mediated thermal conductivity model which calculates the diffusons' contributions to thermal conductivity of amorphous materials[49]. The calculation formula is as follows:

$$\Lambda_{min} = 0.76 k_B n^{2/3} \frac{1}{3}(v_L + 2v_T) \qquad (6)$$

The calculated thermal conductivities of the as-deposited parylene C and the 200 °C-annealed parylene C are both ~0.12 W/m-K, consistent with the measured values, shown in Figure 5. The agreement between the values calculated by the model and the measured thermal conductivities shows that the thermal conduction in the above parylene C samples with a lower thickness (such as 270 nm) is dominated by diffusons.

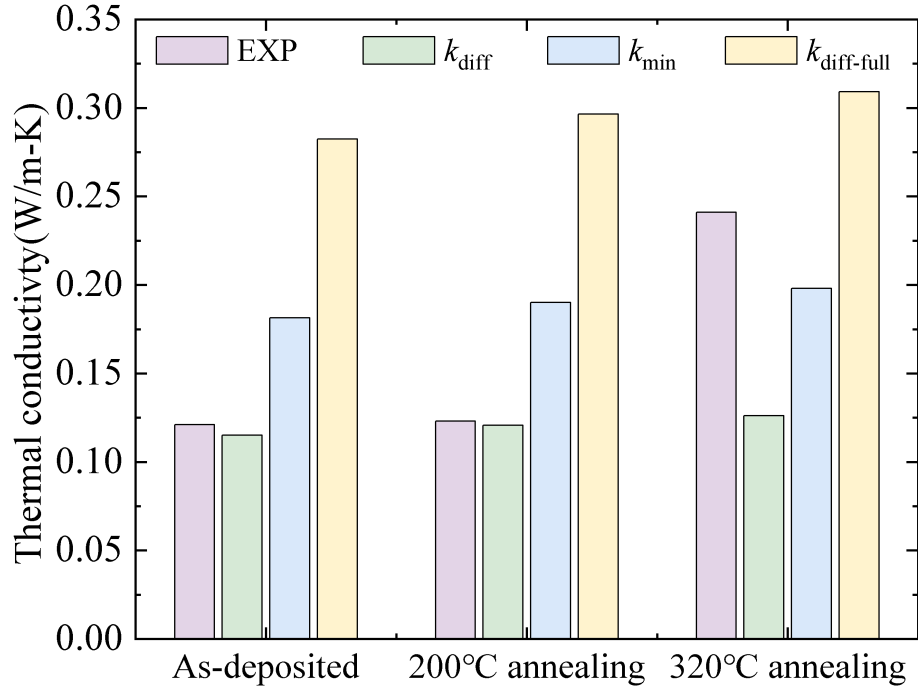

Figure 5. Comparison of the measured and calculated thermal conductivities of the 260-nm-thick as-deposited parylene C thin film, the 257-nm-thick parylene C annealed at 200 °C, and the 160-nm-thick parylene C annealed at 320 °C. EXP is the measured thermal conductivity. $k_{min}$ and $k_{diff}$ are the calculated thermal conductivities by the minimum thermal conductivity model and the diffuson-mediated thermal conductivity model using the effective atomic number, respectively, while $k_{diff\text{-}full}$ is the calculated thermal conductivity by the diffuson-mediated thermal conductivity model using the full atomic number.

Figure 5 compares the measured and calculated thermal conductivities of the three parylene C samples. For the as-deposited and 200 °C-annealed Parylene C, the measured values are consistent with $k_{diff}$, but fall significantly below $k_{diff\text{-}full}$. This confirms that hydrogen atoms are not thermally excited and don't contribute to thermal conductivity. It is evident that $k_{diff}$ is 37% lower than $k_{diff\text{-}full}$.

*3.5. Thermal insulation*

In Figure 6, the thermal conductivity does not strictly correlate positively with their dielectric constant. Notably, among all dense low-k materials, the parylene C exhibits the lowest thermal conductivity, significantly lower than that of traditional inorganic materials such as $SiO_2$ and FSG, while maintaining a relatively low dielectric constant. This means that it can serve not only as a dielectric layer in microelectronics devices[1,50], but also as a thermal insulation layer in the field of integrated circuits, reducing thermal crosstalk between high-power components (logic, CPU) and temperature-sensitive components (memory)[27]. Meanwhile, parylene C has favorable mechanical strength and structural stability, which is particularly crucial for microelectronic devices and MEMS that require thermal insulation while withstanding mechanical stress[7]. Although porous materials may have lower thermal conductivity and dielectric constant, the poor thermal and mechanical stability will increase more issues[51,52]. Therefore, parylene C thin films, as functional coatings, possess distinctive advantages in the fields of advanced packaging, micro-electro-mechanical systems, and thermal insulation due to the low dielectric, the ultralow thermal conductivity, and favorable mechanical strength.

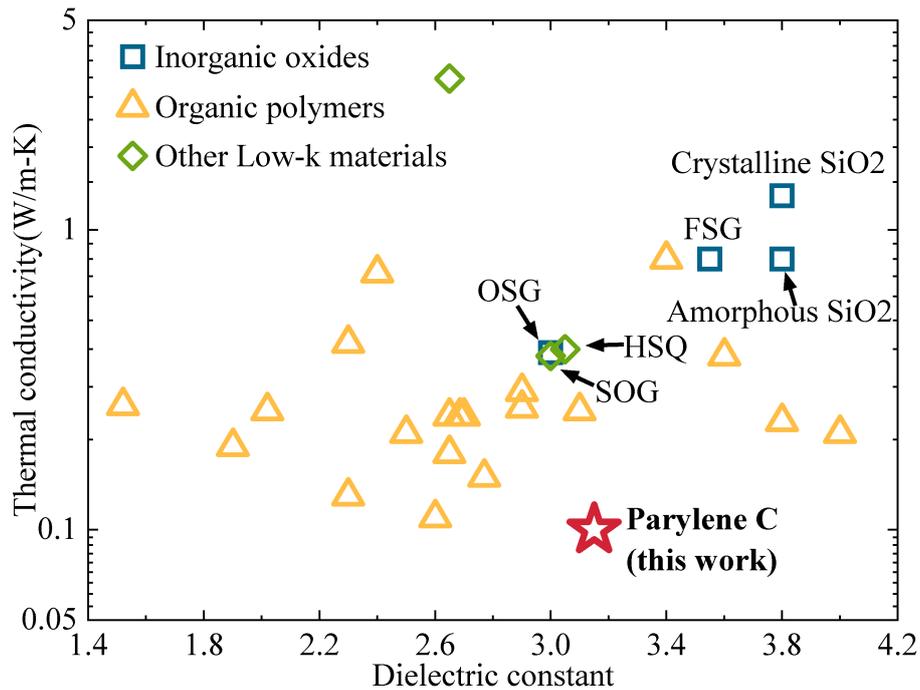

Figure 6. Comparison of the thermal conductivity of parylene C with that of other dense low-dielectric constant materials, including inorganic oxides, organic polymers, and other low-k materials (organic-inorganic hybrid materials and composite). Detailed information can be found in the supplementary information.

## 4. Conclusions

This work discussed the relationship between the structure, regulated by thickness and post-annealing temperature, and the cross-plane thermal conductivities of parylene C thin films at room temperature and under high-temperature conditions. The negligible variations in thermal conductivity of the 200 °C-annealed parylene C confirm that inter-chain *van der Waals* bonds remain the primary bottleneck of heat conduction along the cross-plane direction. The enhancement in thermal conductivity observed in the 320 °C-annealed parylene C results from a synergistic structural optimization: melt-recrystallization simultaneously increases crystalline quality and changes the molecular chain orientations, thereby reducing phonon scattering and establishing efficient cross-plane covalent pathways for thermal conduction. The calculated values from the diffuson-mediated thermal conductivity model are consistent with the measured thermal conductivities indicating that the thermal conductivity is dominated by diffusons for the as-deposited and 200 °C-annealed parylene C films with lower thicknesses. Meanwhile, parylene C possesses the lowest thermal conductivity among dense low-k materials. These results offer a comprehensive understanding of heat transport mechanisms and thermal conductivity modulation methods of linear polymers, supporting the thermal insulation in advanced packaging.

**CRediT authorship contribution statement**

Yicheng Wei: Writing – original draft, Writing – review & editing, Investigation, Data curation, Formal analysis, Visualization. Han Xu: Writing – original draft, Writing – review & editing, Resources, Data curation, Formal analysis, Visualization. Xingqiang Zhang: Software, Data Curation. Wei Wang: Writing – review & editing, Supervision. Zhe Cheng: Writing – review & editing, Conceptualization, Methodology, Validation, Supervision, Funding acquisition.

**Declaration of competing interest**

The authors declare no competing interest.

**Acknowledgments**

This study was supported by the National Key Research and Development Program of China (Grant No. 2024YFA1207900), the National Natural Science Foundation of China (Grant No. 62574007, T2550270).

**Data availability**

The data that support the findings of this study are available from the corresponding authors upon reasonable request.